\documentclass[12pt]{iopart}

\usepackage{cite}

\usepackage{graphicx}
\usepackage{color}
\usepackage{soul}
\usepackage{bm}
\usepackage{iopams}

\expandafter\let\csname equation*\endcsname\relax

\expandafter\let\csname endequation*\endcsname\relax

\usepackage{amsmath}
\usepackage[breaklinks=true,colorlinks=true,linkcolor=blue,urlcolor=blue,citecolor=blue]{hyperref}

\DeclareSymbolFont{usualmathcal}{OMS}{cmsy}{m}{n}
\DeclareSymbolFontAlphabet{\mathcal}{usualmathcal}

\usepackage[T1]{fontenc}

\usepackage{graphicx}
\begin{document}
\title{A random planting model}
\author{J. Talbot}
\address{Sorbonne Universit\'e, CNRS, Laboratoire de Physique Th\'eorique de la Mati\`ere Condens\'ee (UMR CNRS 7600), 4 Place Jussieu, 75005 Paris, France}
\author{P. Viot}
\address{Sorbonne Universit\'e, CNRS, Laboratoire de Physique Th\'eorique de la Mati\`ere Condens\'ee (UMR CNRS 7600), 4 Place Jussieu, 75005 Paris, France}
\author{D. Colliaux}
\address{Sony CSL Paris, France}

\date{\today}

\begin{abstract}

The adoption of agroecological practices will be crucial to address the challenges of climate change and biodiversity loss. Such practices favor the cultivation of plants in complex mixtures with layouts differing from the monoculture approach of conventional agriculture. 
Inspired by random sequential adsorption processes, we propose  a one-dimensional model in which the plants are represented as line segments that start as points and grow at a constant rate until they reach length $\sigma$ after a  time interval $\tau$.    The planting positions and times are randomly chosen with the constraint that plant overlap is forbidden.  We apply an exact, event-driven simulation to investigate the resulting spatiotemporal patterns and yields in both mono- and duocultures.  After a transient period, with oscillations in the density and coverage, the field reaches a steady state in which the mean age of plants is one half of the time to maturity.  The structure of the active plants is characterized by correlation functions between the fluctuation of the age of a plant and its $k$th neighbour. Nearest neighbours are negatively correlated, while next nearest neighbours tend to have similar ages. The steady state yield increases with the planting rate and approaches a maximum value of 4/3 plants per unit length per unit time. For two species with the same size at maturity but different growth rates, the more slowly growing species is enriched in the harvest compared to the seed mix composition. If two species have the same time to maturity but different sizes, the smaller one is enriched in the harvest and, at a sufficiently high planting rate, the larger species may be completely absent. For two species with the same ratio of $\sigma/\tau$ the selectivity is insensitive to the planting rate. 
This model and the algorithms describing the planting strategies may be extended to higher dimensions, more species and other planting strategies that may assist in the design of novel microfarms.\\

\noindent{\bf keywords:} agro-ecosystem, event-driven simulations, plant growth dynamics.

\end{abstract}

\submitto{J. Stat. Mech.}

\maketitle
\section{Introduction}

As the world is facing major challenges related to climate change and biodiversity loss, new practices for farming are increasingly necessary. Agroecology \cite{altieri2018agroecology, wezel2009agroecology} has been proposed as a path towards sustainable food systems \cite{de2011agroecology,food2021long}. This approach considers crops in their ecosystems, including population dynamics aspects as well as interactions between plants, animals species and micro-organisms. It favours highly diverse agroecosystems and biological control of weeds and pests. These practices were originated by Parisian market gardeners in the 19th century who were cultivating as densely as possible on small plots. This French Intensive Method was rediscovered in the 1970s in the US \cite{Orin} and is now regaining popularity in France \cite{herve2016miraculous}. Computational agroecology is the scientific discipline focusing on digital technologies useful in developping such complex agroecosystem \cite{hanappe2016agroecology, raghavan2016computational,colliaux2022computational} and examples of this are in the systematic exploration of plant association \cite{paut2020modelling} or robotic platforms for planting, harvesting and controlling weeds \cite{colliaux2017lettucethink,ditzler2022automating}.  

A hallmark of agroecological practices, intercropping involves growing two or more crops together 
with the goal of producing a greater yield on a given piece of land by making use of resources or ecological processes that would otherwise not be utilized by a single crop. This is quantified by the land equivalent ratio (LER) that describes how well the association performs in terms of yield compared to the separate cultivation of the crops. Recent studies found that a LER greater than unity can be obtained for many intercropping mixtures \cite{li2023productive, Paut2024}.

It is fruitful to investigate intercropping from a theoretical viewpoint \cite{vandermeer1992ecology}. In order to model the process it is necessary to specify the planting positions and growth dynamics of the plants. The plant arrangements can be approximated as a set of disks with various radii \cite{Ecormier-Nocca2019} and various studies have shown that a logisitic, or sigmoidal, growth model provides a good description \cite{Tei1996, thornley2007mathematical}. Previous studies have addressed the spatial dispositions of plants in intercrops and the effect of sowing times \cite{vandermeer1992ecology,Garcia-Barrios2001}. Bimodal size distributions, which result from competition between large and small plants, have been observed in monospecific plant distributions \cite{Ford1975,Huston1986}. 

In this study we propose a random planting model \cite{colliaux2023models}, with a simple geometric representation of the plants and a small set of parameters, that is related to the random sequential adsorption (RSA) process \cite{Talbot2000, Kubala2022}. In the one-dimensional version, also known as the parking lot model  \cite{renyi1958one}, rods of unit length are placed randomly and sequentially on a line without overlap. In the ``jamming'' limit about 75\% of the line is covered by rods. RSA has been extended to bi- and polydisperse populations \cite{TalbotMixtures1989,tarjus1991random,Hassan2002,Wagaskar2020} and to growing particles \cite{Boyer1994,Dodds2002,Andrienko1994}.  
RSA is an irreversible process: once an object has been placed at a certain location it remains there indefinitely. A consequence is that,  particularly at high density, RSA configurations differ significantly from the corresponding equilibrium systems \cite{Widom1966,Torquato2006}. When a relaxation mechanism, such as desorption, is present the system evolves to an equilibrium state\cite{Oleyar2007,Talbot2000b}. It is this reversible parking lot model that is most appropriate for the present application. In the 1d case, in the limit of weak desorption the close-packed state (coverage unity) is approached slowly in time,  $1/\log(t)$ \cite{Talbot2000b}. The model  can describe the compaction of granular matter and the reversible adsorption of proteins on a solid surface.

In our model, a new plant is placed in the field a random position with the condition that it will never overlap with any other active plant. Each plant is harvested after a fixed time. Although specific properties of intercropped systems arise due to interactions among plants, here we investigate the properties of intercropped systems without these interactions. We assume linear growth model which, while less realistic than the sigmoidal growth curve, has the advantage of allowing us to thoroughly explore the properties of the model.  Although analytic solutions of the 1D RSA model and the corresponding equilibrium system of hard rods on a line have been obtained, it appears to be considerably more  difficult to do the same with this random planting model. Instead, we have developed an efficient event-driven simulation algorithm that we have used to study the kinetic and steady state properties of the model. In particular, we are interested in the yield (number of plants harvested per unit time, per unit length) as a function of the planting rate and, in the case of two species, the fraction of each in the harvest as a function of the seed mixture composition. The selectivities can be compared with those of related systems \cite{Talbot1997}. We also examine the spatio-temporal correlations of plants in the field.

\begin{figure}
 \begin{center}
  \includegraphics[scale=0.4]{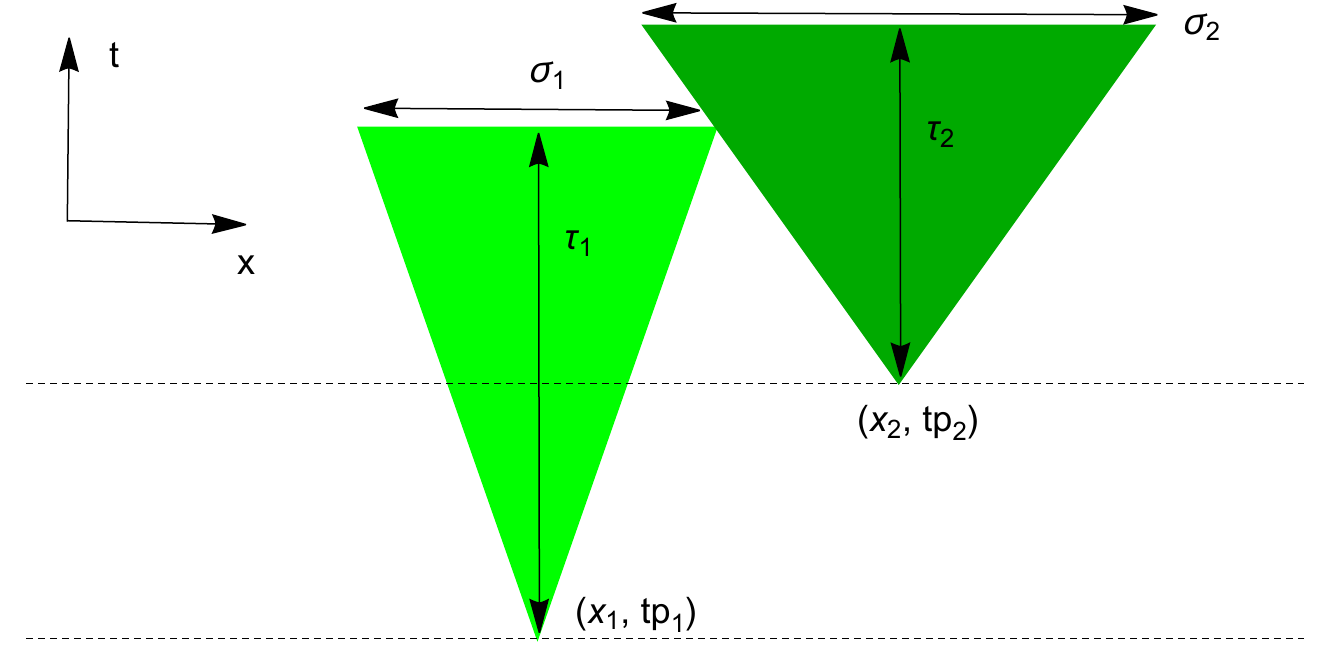}
 \end{center}
   \caption{Space-time diagram of the model. The left plant was planted as a grain of size zero at time $t_{p_1}$ and position $x_1$. 
   It grows linearly in time until it reaches a size $\sigma_1$ at which point it is harvested. Plant 2 was planted afterwards at time  $t_{p_2}$ and position $x_2$. It comes into contact with plant 1 just as the latter is harvested, then continues to grow until reaching maturity with size $\sigma_2$. The positions of new plants are chosen so that they never overlap with any active plant.}
   \label{fig:model}
 \end{figure}

\section{Model}
We assume that a new plant starts as a grain that initially occupies zero length and then grows linearly in time until it reaches a certain size at which point it is harvested. A species $i$ is characterized by its maximum diameter, $\sigma_i$ and the time taken to grow to this size, $\tau_i$. The first plant is planted at $t=t_{p_1}=0$ at a position chosen from a uniform distribution in the interval $(0,L)$. 

The idea is to sow new seeds that will develop into plants that will never overlap with others that are already growing in the field.  
Let us assume that plant 2 is planted after plant 1. Clearly no overlap can occur if $t_{p_2}-t_{p_1}>\tau_1$, as the second plant is only introduced after the first has been harvested. If $t_{p_2}-t_{p_1}<\tau_1$ there are two possibilities:  If $t_{p_2}-t_{p_1}<\tau_1-\tau_2$ then the two triangles (plants) overlap if
\begin{equation}
|x_2-x_1| <\frac{\sigma_2}{2}+\frac{\sigma_1}{2\tau_1}(t_{p_2}-t_{p_1}+\tau_2)
\end{equation}
while if $t_{p_2}-t_{p_1}>\tau_1-\tau_2$ the overlap condition is
\begin{equation}
|x_2-x_1| <\frac{\sigma_1}{2}+\frac{\sigma_2}{2\tau_2}(t_{p_1}-t_{p_2}+\tau_1)
\end{equation}
If no overlap is detected, the new plant is accepted and its parameters $t_{p_i}, \sigma_i,\tau_i$ are recorded.

\section{Event driven simulation}

A naive algorithm to simulate this model is as follows: Attempts to sow a new plant are made at a rate $k_p$.  A species is selected with a fixed probability and a position on the line is selected from a uniform random distribution. If, in this position, the selected plant can reach maturity without ever overlapping with its neighbours, it is accepted. If not, the attempt is abandoned. This algorithm has the advantage that it is simple to program and is fairly efficient as long as the coverage of the field is not too high. At high coverages, however, most attempts to add a new plant result in failure. The computation therefore becomes inefficient, requiring the detection of many overlaps before a successful insertion is possible.

We have therefore developed a fully event driven simulation for both the single and two-species models using the Gillespie stochastic algorithm \cite{Gillespie1977,Rao2003,TV06}. Instead of a fixed timestep, the method is based on the event rate for a given configuration. The waiting time between successive events is sampled from an exponential distribution. All selected events are implemented, thus avoiding the need to make trial moves. This results in significant improvements in efficiency, particularly at high densities, compared with the naive algorithm. Usually in event driven simulations all events occur after an exponentially distributed waiting time. In the present model the plants are harvested after a fixed time, which depends on the species. Fortunately this requires only a minor modification of the algorithm as detailed below.

\subsection{Single species}

The field is modelled as a line of length $L$ on which the plants are sowed at randomly selected positions. A plant configuration consists of a number of plants in various stages of development. The planting positions are chosen in such a way as to guarantee that no overlap ever occurs. 

Let us introduce the available length, $L_0=L\phi(t)$, which is the length available for the insertion of a new plant so that it never overlaps with any of the previously planted neighbours during its development. Ignoring the harvesting process for the moment, the total event rate (of plantings) is
\begin{equation}
R(t) = k_p\phi(t)
\end{equation}
Since this is a random process the waiting time distribution to the next event is exponentially distributed. We sample the distribution using
$\Delta t = -\log\xi/R(t)$ where $0<\xi<1$ is a uniformly distributed random variate.

Let $\mathcal{A}=\{\mathcal{P}_1,\cdots,\mathcal{P}_m\}$ denote the set of active plants. 
The configuration of the field at time $t$ consisting of $m$ plants in various stages of development is fully characterized by $\{\{x_1,t_{p_1}\},\{x_{2},t_{p_2}\},\cdots,\{x_m,t_{pm}\}\}$ with $x_{1}<x_{2}<\cdots<x_{m}$ and $0\le t-t_{p_k}<\tau_k,\;1\le k\le m$. We will be interested in the density of plants
\begin{equation}
\rho(t)=m(t)/L,
\end{equation}
and the coverage
\begin{equation}
\theta(t)=\sum\sigma_i(t)/L,
\end{equation}
with
\begin{equation}
\sigma_i(t) = \frac{(t-t_{p_i})\sigma_i}{\tau_i}\Theta(t_{p_i}+\tau_i-t),\;t\ge t_{p_i},
\end{equation}
where $\Theta(x)$ is the Heaviside function.

We examine the field and the harvesting times of all the plants that are growing. Let $t_{min}=\min_{k}(t_{p_k}+\tau_k)$ for $k\in \cal{A}$ denote the time to the next harvest. There are two possibilities: 
(i) $t+\Delta t < t_{min}$, i.e. the next event (sowing a new plant) occurs before the first harvest;  (ii) $t+\Delta t > t_{min}$. 

In case (i) the system time is advanced to $t'=t+\Delta t$ and a new plant is sowed uniformly in the available length.  In case (ii) the system time is advanced to $t'=t+t_{min}$ and the next plant to be harvested is removed from the list of active plants.

The simulation parameters are $L/\sigma$, $k_p$ and $ED$, the event density per unit length.

\subsection{Two species}

\begin{figure}[t]
	\begin{center}
		\includegraphics[width=13cm]{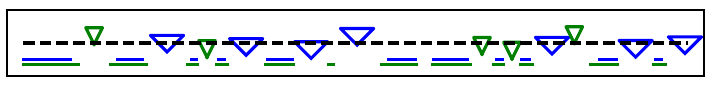}
	\end{center}
	\caption{Space-time diagram of the field for two species showing the active plants at the current time (dashed line). The solid lines (displaced from the dashed line for clarity) show the free lengths for the insertion of the blue and green species. The tip of a new triangle will be inserted uniformly in the available length. Note that some segments between two active plants are accessible only to the smaller green species.}
	\label{fig:FreeLengths}
\end{figure}

With two species the simulation algorithm is similar to the single species case but with the following modifications.
Let us introduce the insertion probabilities, $\phi_1(t)$ and $\phi_2(t)$ of the green and blue species, where $L\phi_1(t)$, for example, is the length available for the insertion of the center of a green plant so that it never overlaps with any of the previously planted neighbours during its development. See Fig.\ref{fig:FreeLengths}. The total event rate (of plantings) is
\begin{equation}
R(t) = k_p[x_1\phi_1(t)+x_2\phi_2(t)]
\end{equation}
where $x_1$ and $x_2$ are the fraction of green and blue seeds, respectively ($x_1+x_2=1$).

At the next event, a plant of species $k$ is selected with probability $k_p x_k\phi_k(t)/R(t)$. Note that the calculation of the available segments is restricted to the nearest neighbours only:  if the species differ too much in size some overlaps may be neglected. 

\section{Monoculture}

We first consider the cultivation of a single species with parameters $\sigma$ and $\tau$ and discuss the maximum yield that can be obtained in a non-random, deterministic case.
\subsection{Regular planting}

\begin{figure}
	\begin{center}
		\includegraphics[width=8cm]{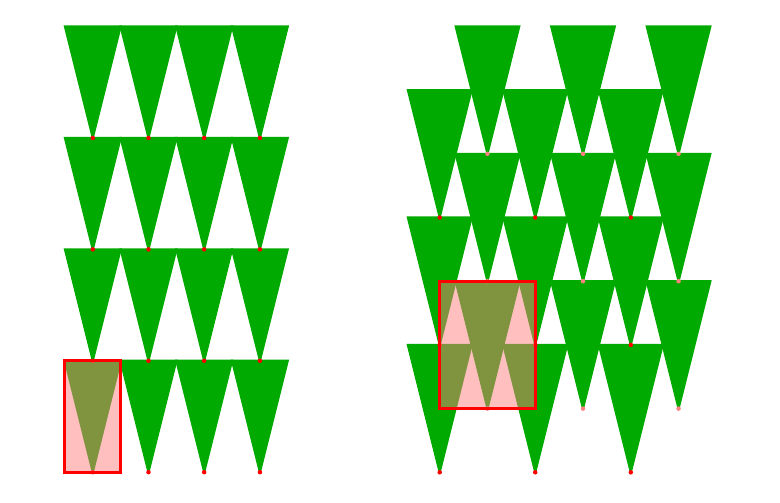}
	\end{center}
	\caption{Idealized crop rotations for a single species characterized by parameters $\sigma, \tau$. Left: Synchronised: Four plants are sowed with spacing $\sigma$ at the same time. After a time $\tau$ all plants reach maturity and are harvested. A new row is then planted in the same positions.  This scheme results in a harvesting rate of $(\sigma\tau)^{-1}$. Right: Desynchronised: In  the first rotation, three plants are planted with spacings $3\sigma/2$. After a time $\tau/2$, i.e., before the previously sowed plants have reached maturity, a second row of three plants is sowed with the same spacing of  $3\sigma/2$, but offset by $3\sigma/4$. This scheme results in a harvesting rate of $(4/3)(\sigma\tau)^{-1}$. The rectangles show the unit cells of the lattices.}
	\label{fig:Lattices}
\end{figure}

A simple crop rotation plan is shown in Fig. \ref{fig:Lattices}a.  Plants are planted with a spacing equal to the maximum size $\sigma$ and harvested after a time $\tau$. Then new plants are sowed in the same positions. The harvesting rate is simply $1/{\sigma\tau}$ plants per unit time per unit length, which corresponds to the density of the triangles in the space-time diagram. This method appears to be efficient because there is no waiting time is introduced and the density seems to be maximum. 

However, the yield can be increased with desynchronised planting, in which new plants are inserted before their neighbours reach maturity. In the example shown in \ref{fig:Lattices}b the first rotation is planted with a spacing of $3\sigma/2$. After a time $\tau/2$ a second group of plants is inserted midway between those in the first layer. The latter are harvested just as they come into contact with the second layer plants.  This results in a unit cell of the lattice of dimensions $(3\sigma/2,\tau)$  that produces two plants for a yield of $4/(3\sigma\tau)$ plants per unit time per unit length.  It turns out that this is the maximum. See \cite{talbot2024optimizing} for more details, as well as extensions to two-dimensions and sigmoidal growth curves. In this idealized comparison, desynchronisation results in a 33\% increase in efficiency. It can be shown that this is a maximum and extended to two-dimensions. 

\begin{figure*}[t]
	\begin{center}
		\includegraphics[width=13cm]{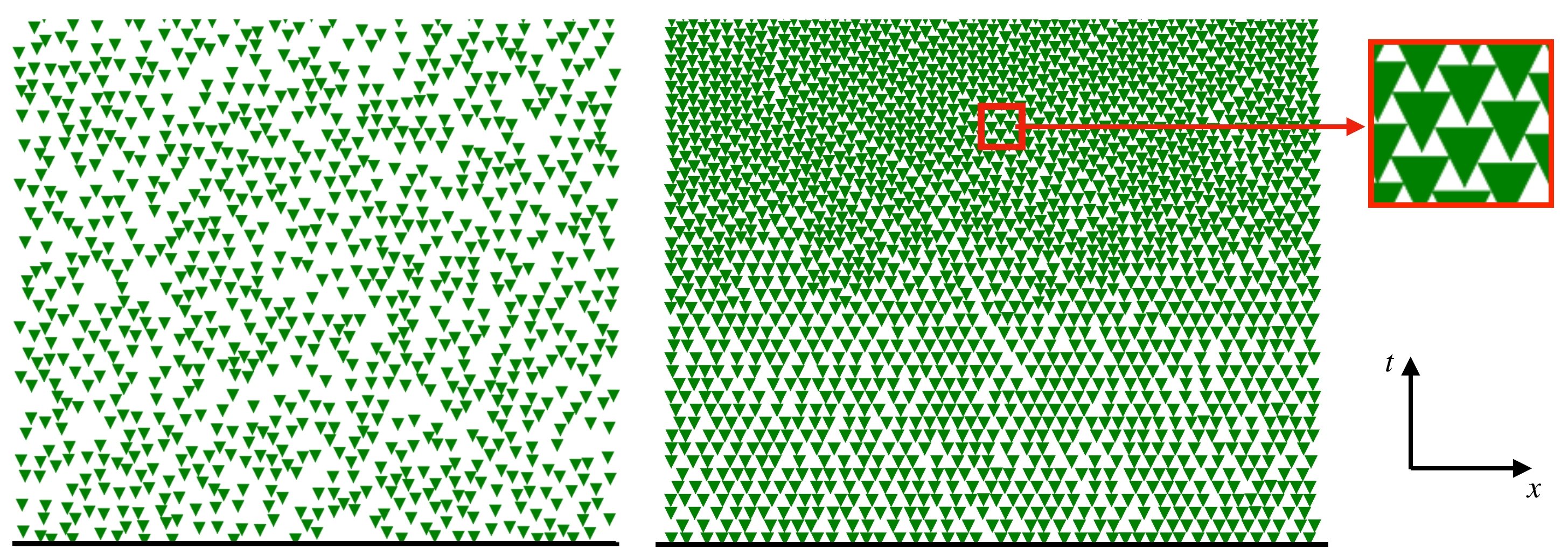}
	\end{center}
	\caption{Space-time diagram of a monoculture event driven simulation for  $k_p=1 ({\rm left}),1000$. A steady state is rapidly obtained in the first case, but much more slowly for $k_p=1000$.}
	\label{fig:Fields}
\end{figure*}

\subsection{Random planting}
We now consider the random planting model.
For all planting rates, the dynamics of the process always evolves towards a steady state,  after a typical relaxation time. 
 
Figure \ref{fig:Fields}  shows the evolution of the field in a space-time diagram for a monoculture with planting rates of $k_p=1,1000$. At large $k_p$, since the interval between two planting events is so small, the behaviour is initially very close to that of a simple RSA process of rods of length $\sigma$. After time $\tau$ these plants reach maturity and are harvested. A new "layer" is then added and this cycle continues until slight differences in the planting times lead to desynchronization and densification of the field. The system eventually reaches a steady state and no further net evolution of the structure is observed,  although fluctuations persist. 
 
\begin{figure}
\begin{center}
 	\includegraphics[width=6cm]{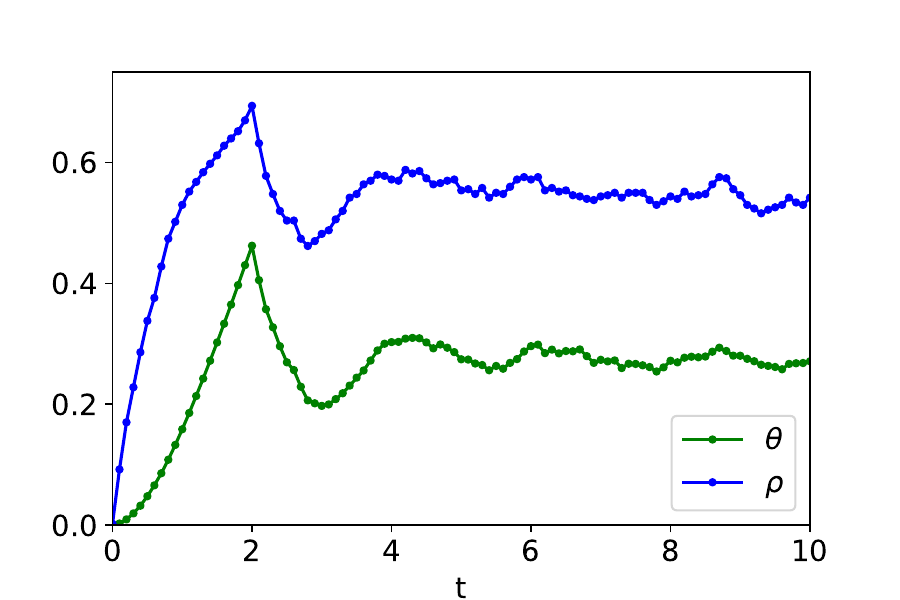}   
 \includegraphics[width=6cm]{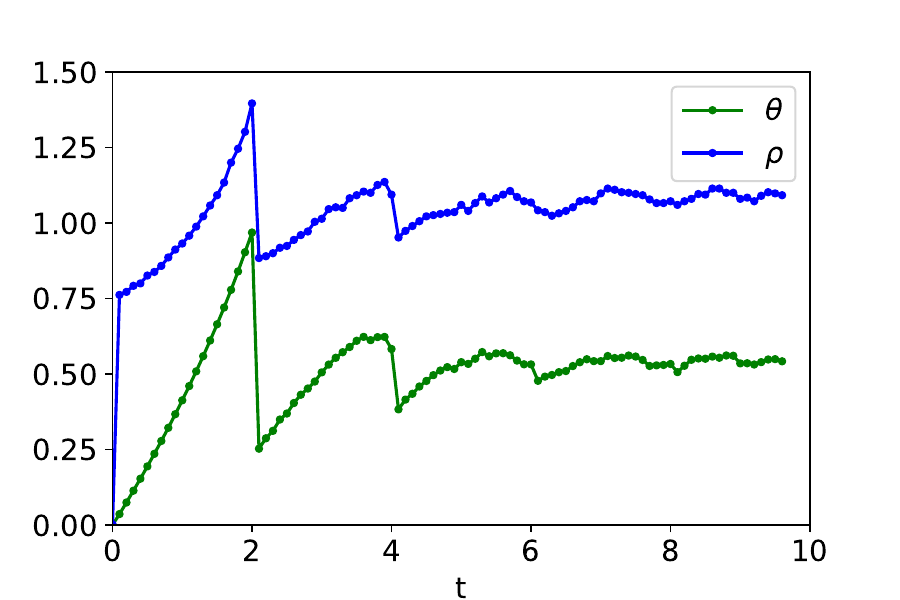}    
\end{center}
\caption{Kinetics of a monoculture on a field of length $L=500$ with $\tau=2,\sigma=1$. The green and blue lines show the coverage and the density, respectively. Left: $k_p=1$, Right: $k_p=500$.}

\label{fig:KineticsMC}
\end{figure}

The time evolution of the density and coverage is shown in Fig. \ref{fig:KineticsMC}. For high planting rates the density increases rapidly to a threshold value, while the evolution of the coverage is limited by the growth rate, $\sigma/\tau$, of the plants. After the initial rise, one observes oscillations in both quantities at short times. This results from the near simultaneous planting of many plants at the beginning of the simulation that are harvested at almost the same time resulting in a sudden drop in the density and coverage. After a number of such cycles, the planting  becomes desynchronized and the quantities fluctuate around their steady state values. 

\begin{figure}
 \begin{center}
         \includegraphics[width=6.2cm]{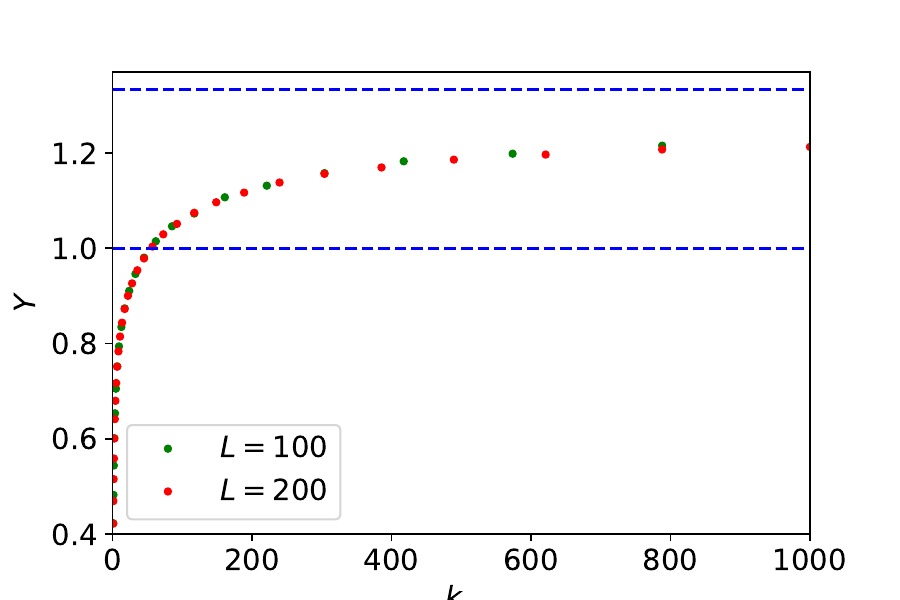} 
        \includegraphics[width=6.2cm]{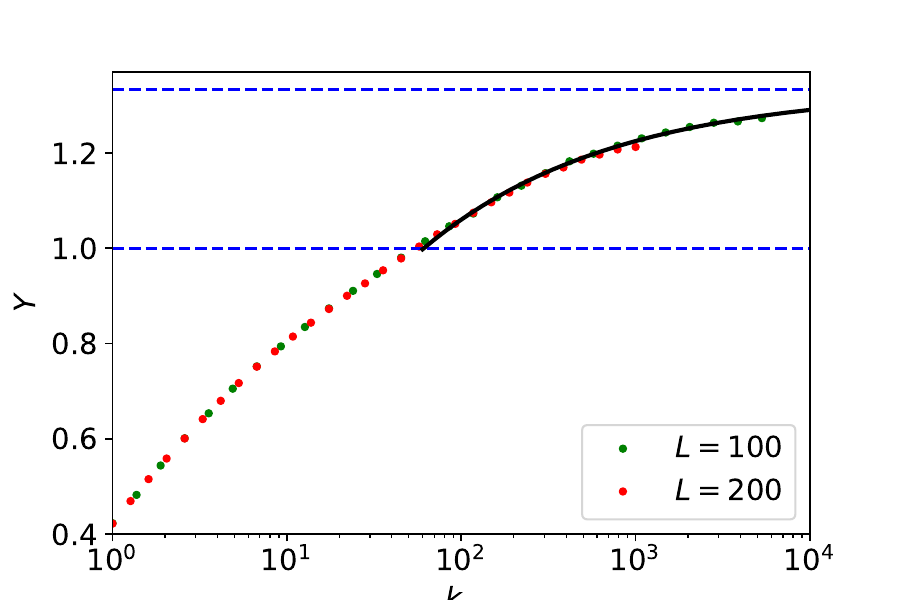} 
        \end{center}
   \caption{Left: Yield versus the planting rate in the steady state for a monoculture for fields of lengths $L=100,200$; Right: Same with a log scale for $k_p$. The dashed lines show value of 1 and 4/3 and the solid line shows a fit of $Y=4/3-a/k_p^b$. }
    \label{fig:YieldVsKp}
\end{figure}

Figure \ref{fig:YieldVsKp} shows that the steady state density, or yield, is a monotonically increasing function of the planting rate. It appears to approach the value of the optimum regular, desynchronised planting, namely $4/3$. The yield as a function of the planting rate can be described by the empirical equation $Y=4/3-a/k_p^b$ with $b\approx 0.4$.



We now focus on  the statistics of the ages of the active plants, which can serve as order parameters. The age of active plant $i$ at time $t$ is
\begin{equation}
a_i=t-t_{pi}
\end{equation}
from the properties of the model we have $0\le a_i\le \tau$. If the ages are uniformly distributed between these two limits we have
\begin{equation}
\langle a_i\rangle = \tau/2,\;\;\langle a_i^2\rangle = \tau^2/3
\end{equation}

\begin{figure}
 \begin{center}
          \includegraphics[width=6.2cm]{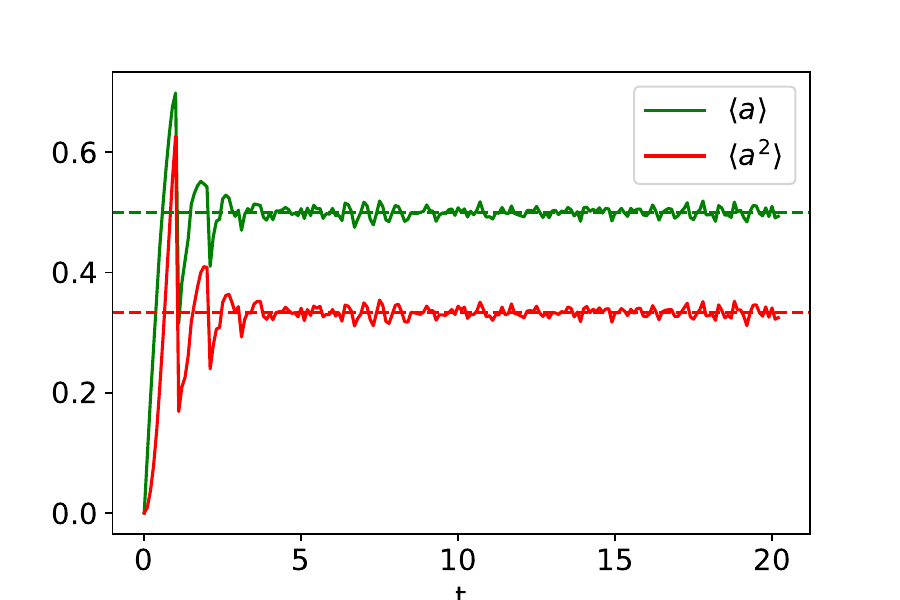} 
        \includegraphics[width=6.2cm]{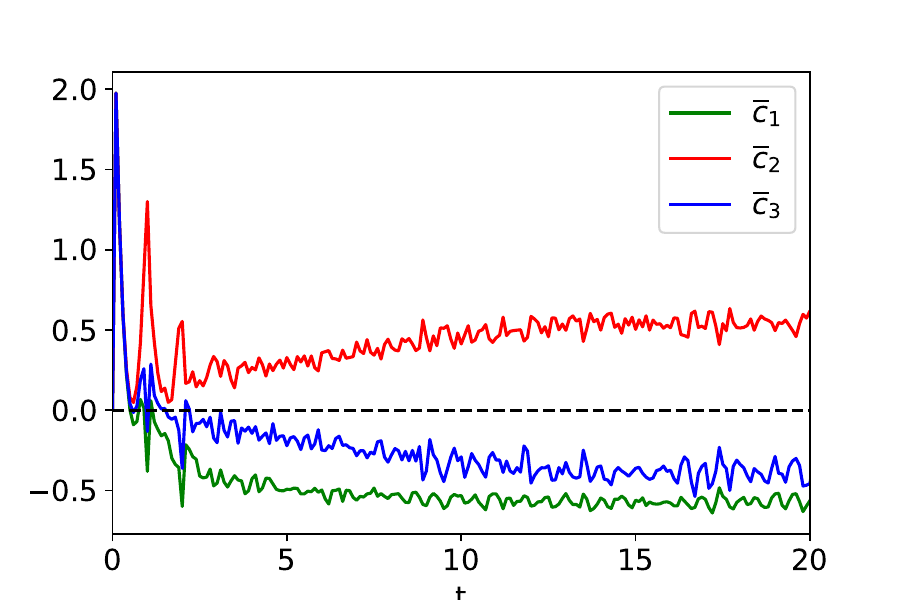} 
 \end{center}
  
   \caption{Left: Mean (green) and mean squared (red) ages of active plants. Right: Evolution of the correlation functions $\bar{c}_1,\bar{c}_2,\bar{c}_3$. In both cases, $\tau=1,\sigma=1,L=200,k_p=500$.}
    \label{fig:AgesCF}
\end{figure}

Discarding the short-time dynamics corresponding to the relaxation time, the steady state values obtained in simulations match the mean age values obtained with plants uniformly distributed. See Fig. \ref{fig:AgesCF}. Even the random planting model to increase the yield for sufficiently large $k_p$, harvesting is now a stationary quantity with fluctuations due to the finite size of the field.
 
We also examine the correlation between the ages of neighbouring plants.
We define the correlation function
\begin{equation}
c_{i,k} = \frac{12}{\tau^2}\langle (a_{i}-\langle a\rangle)(a_{i+k}-\langle a\rangle)\rangle
\end{equation}
which is equal to zero if there is no correlation. For simplicity we examine the average values $\bar{c}_k=\langle c_{i,k}\rangle_i$. 
At low planting rate there is little correlation. For $k_p=500$ the nearest and third nearest neighbours ($k=1,3$) are anti-correlated, while the second nearest neighbours are positively correlated. Increasing the planting rate increases the degree of correlation. The correlations reach the steady state values at a slower rate than the ages. See Fig. \ref{fig:AgesCF}.

 \section{Duoculture}
 
 \begin{figure}[h]
 	\begin{center}
 		\includegraphics[width=8cm]{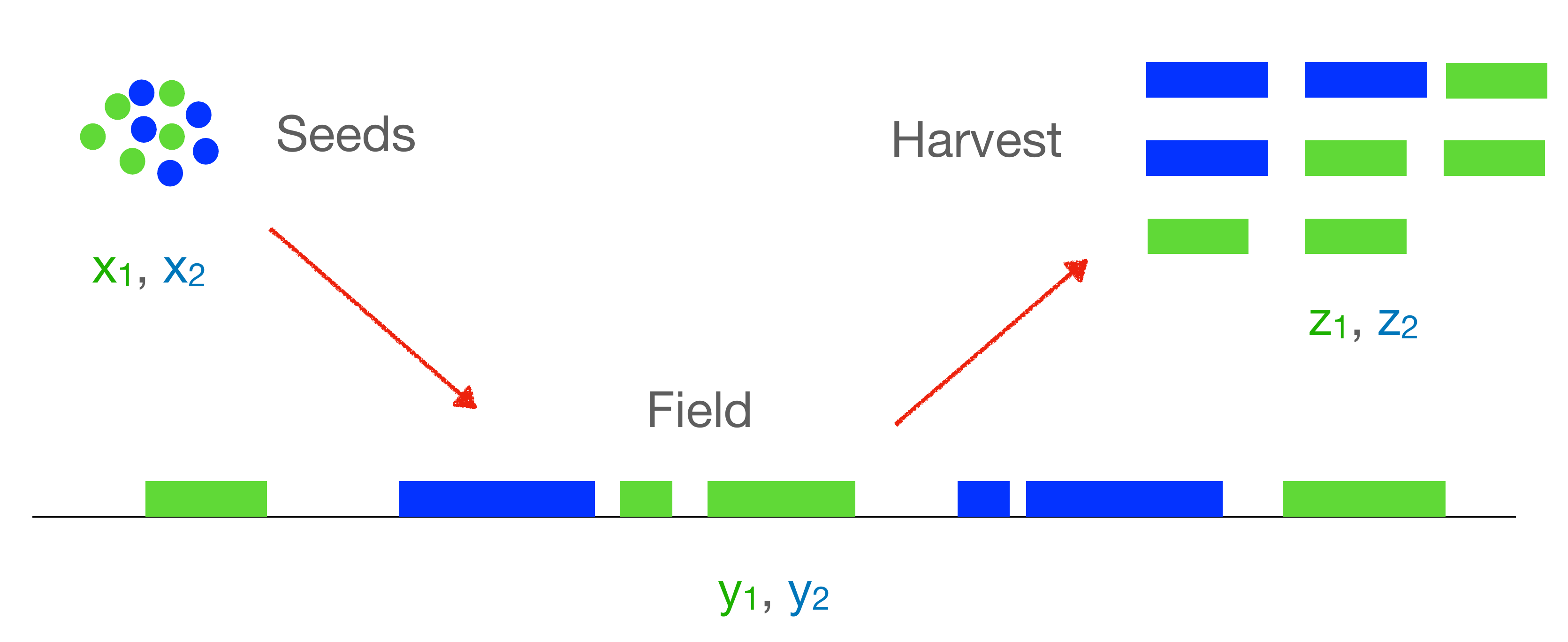}
 	\end{center}
 	\caption{In a duoculture, the seed mix has composition $(x_1,x_2)$. In the steady-state the compositions of the field and the harvested plants are $(y_1,y_2)$ and $(z_1,z_2)$, respectively.}
 	\label{fig:Composition}
 \end{figure}
 With $S$ species present we need specify the seed mixture composition $\{x_j,j=1,S\}$ and then simulate the evolution of the field. We use  $\{y_j,j=1,S\}$ to denote the composition of the field in the steady state and $\{z_j, j=1,S\}$ to denote the composition of the harvested plants. See Fig. \ref{fig:Composition}. These variables satisfy,  e.g., $\sum_1^S x_j=1$. Using the following argument we can relate $\{y_j\}$ and $\{z_j\}$ in the steady-state. In the steady-state species $j$ is present with density $\rho_j$. Each plant of this species reaches maturity at a rate $1/\tau_j$. Therefore the fraction of this species in the harvest is
 \begin{equation}
z_j=\frac{\rho_j/\tau_j}{\sum\rho_k/\tau_k}=\frac{y_j/\tau_j}{\sum y_k/\tau_k}
\label{eq:harvestcomp}
\end{equation}

In this work we have considered the following binary mixtures: 

\begin{center}
\begin{tabular}{|c|c|c|c|c|} 
\hline
mixture & $\sigma_1$ & $\sigma_2$ & $\tau_1$ & $\tau_2$ \\
\hline
M1 & 1& 1 & 1 & 2 \\ 
M2 & 1 & 2 & 1 & 1\\ 
M3 & 1& $a$ & 1 & $a$ \\ 
\hline
\end{tabular}
\end{center}
with $1\le a \le 2$.

With two species of plants the model displays a much richer phenomenology than when only one species is present. We focus on the stationary properties of the model.

\begin{figure}
 \begin{center}
  \includegraphics[width=6cm]{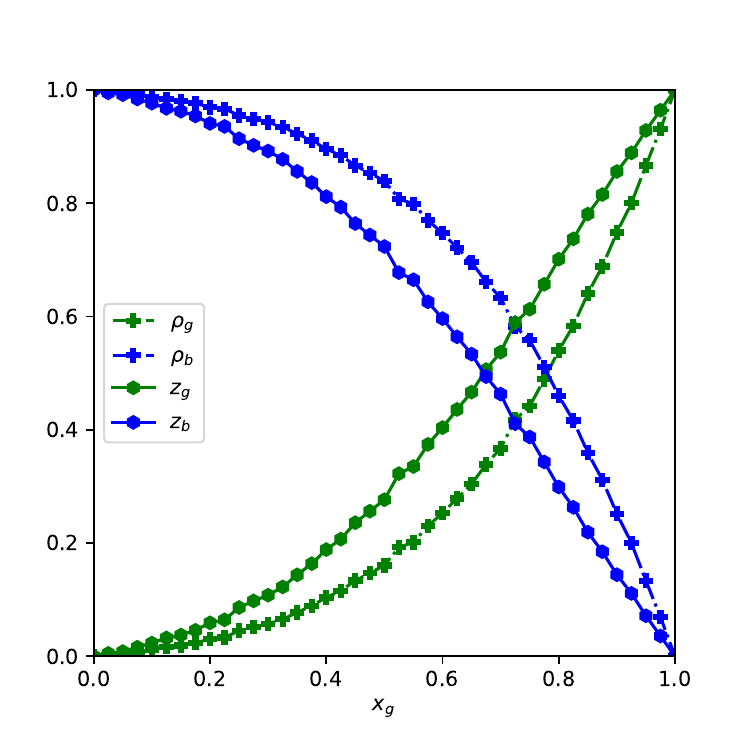}
  \includegraphics[width=6cm]{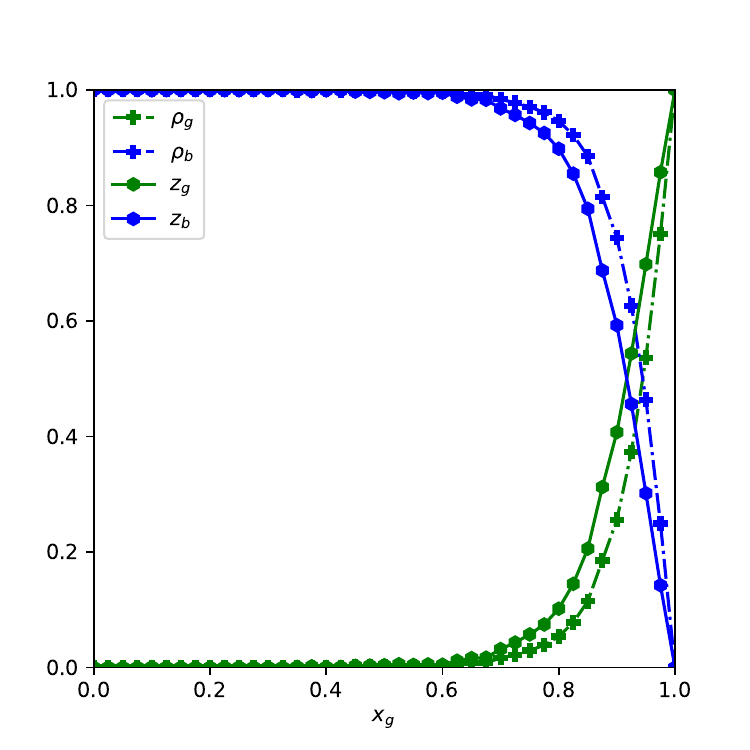}
 \end{center}
   \caption{Steady state composition of the field  $(y_g,y_b)$ (dashed lines) and the harvested plants, $(z_g,z_b)$ (solid lines), as a function of the seed mixture composition, $x_1$, for mixture M1 with $k_p=50$ (left) and $k_p=500$. }
   \label{fig:SteadyM1}
 \end{figure}
\begin{figure}
 \begin{center}
  \includegraphics[width=6cm]{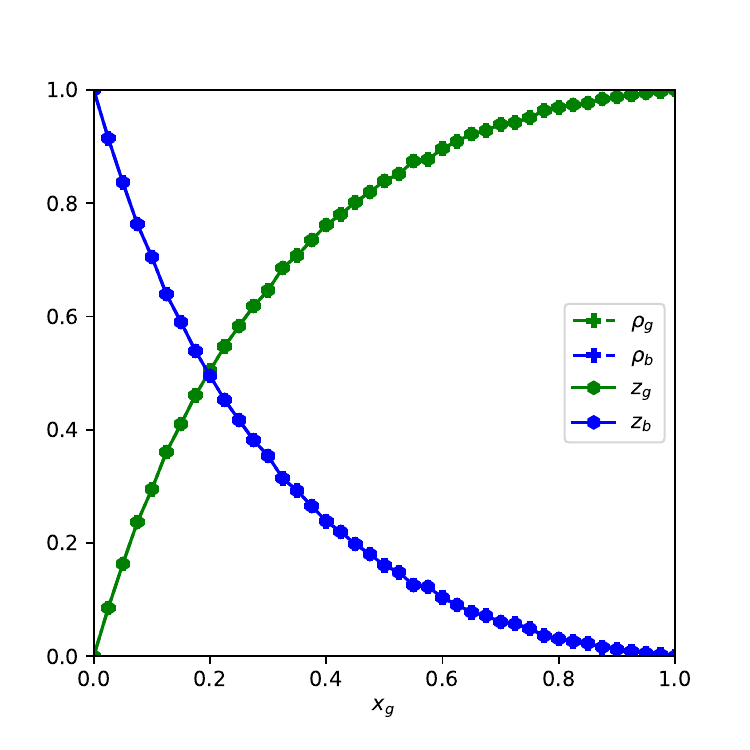}
  \includegraphics[width=6cm]{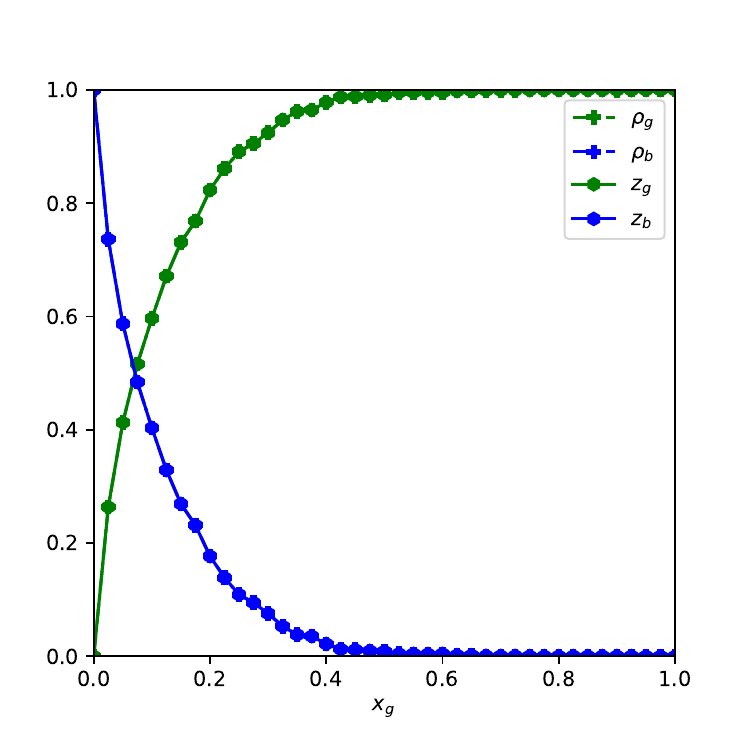}
    
 \end{center}
   \caption{Steady state composition of the field  $(y_1,y_2)$ (dashed lines) and the harvested plants, $(z_1,z_2)$ (solid lines), as a function of the seed mixture composition, $x_1$, for mixture M2 ($\sigma_b=2\sigma_g$) with $k_p=50$ (left) and $k_p=500$ (right).}
   \label{fig:SteadyM2}
 \end{figure}
\begin{figure}
 \begin{center}
\includegraphics[width=6cm]{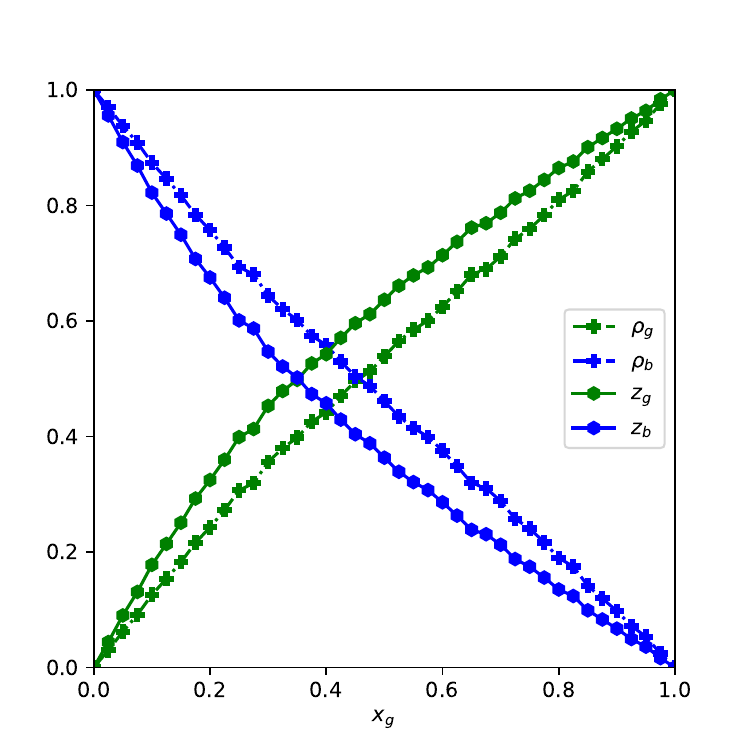}
  \includegraphics[width=6cm]{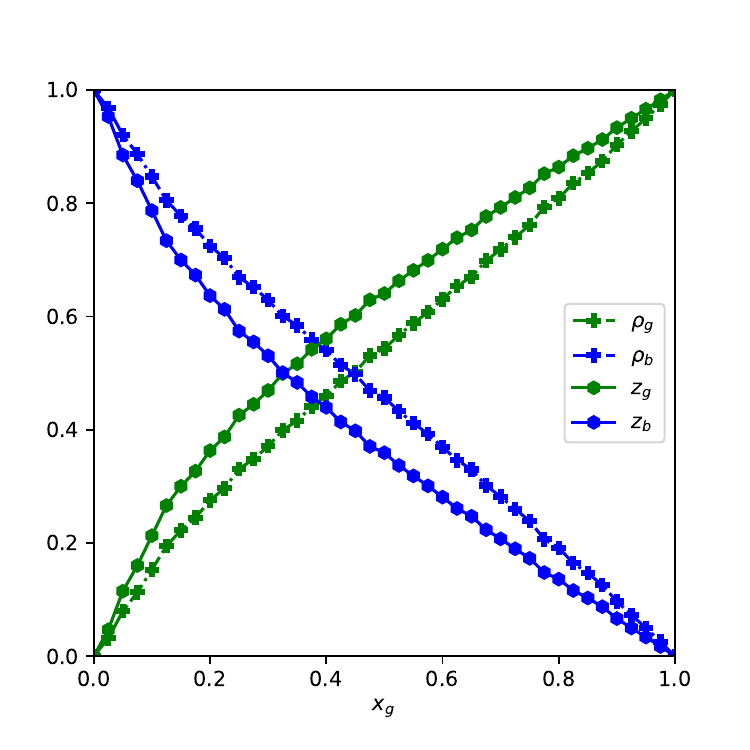}   
   \caption{Steady state composition of the field  $(y_1,y_2)$ (dashed lines) and the harvested plants, $(z_1,z_2)$ (solid lines), as a function of the seed mixture composition, $x_1$, for mixture M3 (with $a=1.5$) with $k_p=50$ (top) and $k_p=500$. }
   \label{fig:SteadyM3}
   \end{center}
 \end{figure}

Figs. \ref{fig:SteadyM1},\ref{fig:SteadyM2},\ref{fig:SteadyM3}, show the composition of the harvested plants as a function of the seed mixture composition, $x_1$ in the steady state. The dashed lines show the 
densities of the active green and blue plants, $\rho_1,\rho_2$, while the solid lines show the composition of the harvested plants, $(z_1,z_2)$. These diagrams can also be used to determine the seed mixture composition necessary to obtain a desired harvest composition. For example for mixture M1 with a planting rate $k_p=50$ to obtain 50\% green plants one would need to plant 65\% green seeds.

{\bf Mixture M1:} Both species have the same size at maturity $\sigma_2=\sigma_1=1$, but the blue plant is allowed to grow for twice as long as the green one before it is harvested, $\tau_2=2\tau_1=2$.  The available line for the deposition of a new plant is the same for the two species, but because the lifetime of the blue plant is twice that of the green plant, the former is favoured in the steady state field. 
The more slowly growing blue plants are enriched in the harvest compared to their presence in the seed mixture. Increasing the planting rate enhances this effect and only blue plants are present in the steady state field for $x_1<0.3 (k_p=50)$ and $x_1<0.8 (k_p=500)$.

{\bf Mixture M2:} The two species are harvested after the same time, $\tau_2=\tau_1=1$, but the size at maturity of the blue species is twice that of the green one, $\sigma_2=2\sigma_1=1$. Figure \ref{fig:SteadyM2} shows the stationary densities and yields of the two species as a function of the seed mixture composition $x_1$. Note that because the lifetimes are the same, the harvest and field compositions are, from Eq. (\ref{eq:harvestcomp}), equal. The smaller plants are enriched in the harvested plants compared to their presence in the seed mixture. At small $x_1$, the blue plants dominate the harvest, but a crossover occurs at a mixture composition less than $0.5$. This effect becomes more pronounced as $k_p$ increases and for $k_p=500$ and $x_1>0.5$ no blue plants are present in the steady state harvest.  Let us recall that there is no interaction between the two species and this effect is the consequence of exclusion effect, which leads to an available line for the blue plants much larger than for the green plants.

{\bf Mixture M3:} consists of two species with the same profile in the space-time diagram, $\sigma_2/\sigma_1=\tau_2/\tau_1=a$.  The larger plant is enriched in the harvest for small values of $x_1$, while for larger values the harvest has almost the same composition as the seed mixture. Moreover, increasing the planting rate has little effect and both species are present in the harvest for all seed mixture compositions.

One can interpret this behavior by noting that the opposite effects observed with mixtures M1 and M2 are almost compensated. However, one observed that for small values of $x_1$, the yield of blue plants drops more rapidly when the size (but keeping the same ratio size over time) of blue plants increases.
 
\begin{figure}
\begin{center}
  \includegraphics[width=6cm]{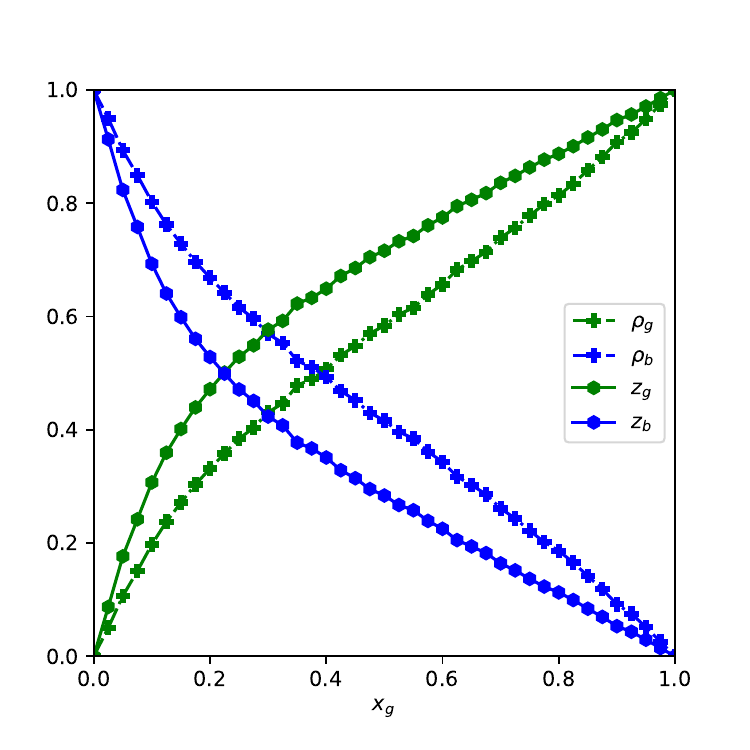}
  \includegraphics[width=6cm]{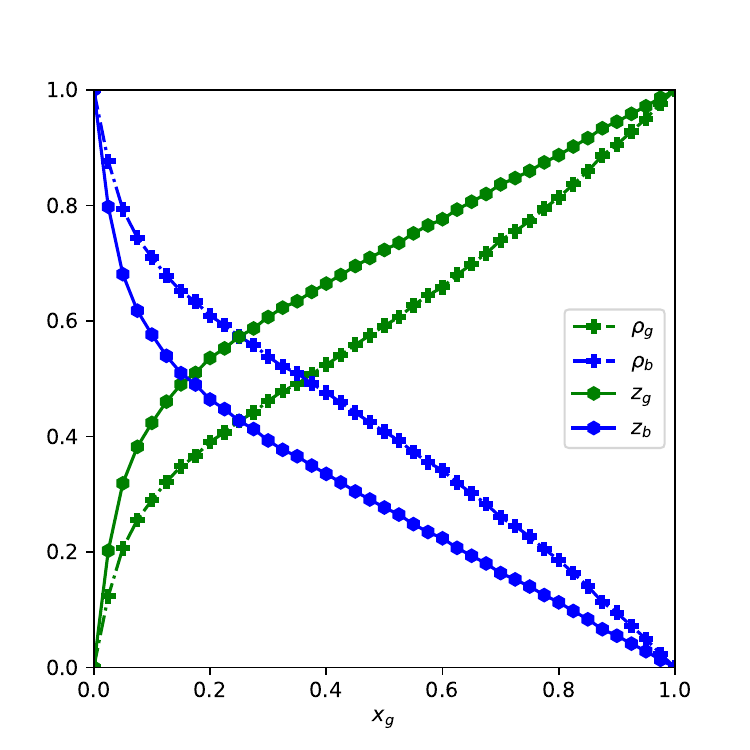}
   \caption{Steady state composition of the field  $(y_1,y_2)$ (dashed lines) and the harvested plants, $(z_1,z_2)$ (solid lines), as a function of the seed mixture composition, $x_1$, for mixture M3 (with $a=1.8$) with $k_p=50$ (left) and $k_p=500$ (right). }
   \label{fig:SteadyM3a}
   \end{center}
 \end{figure}

 \begin{figure*}
 \begin{center}
  \includegraphics[width=18cm]{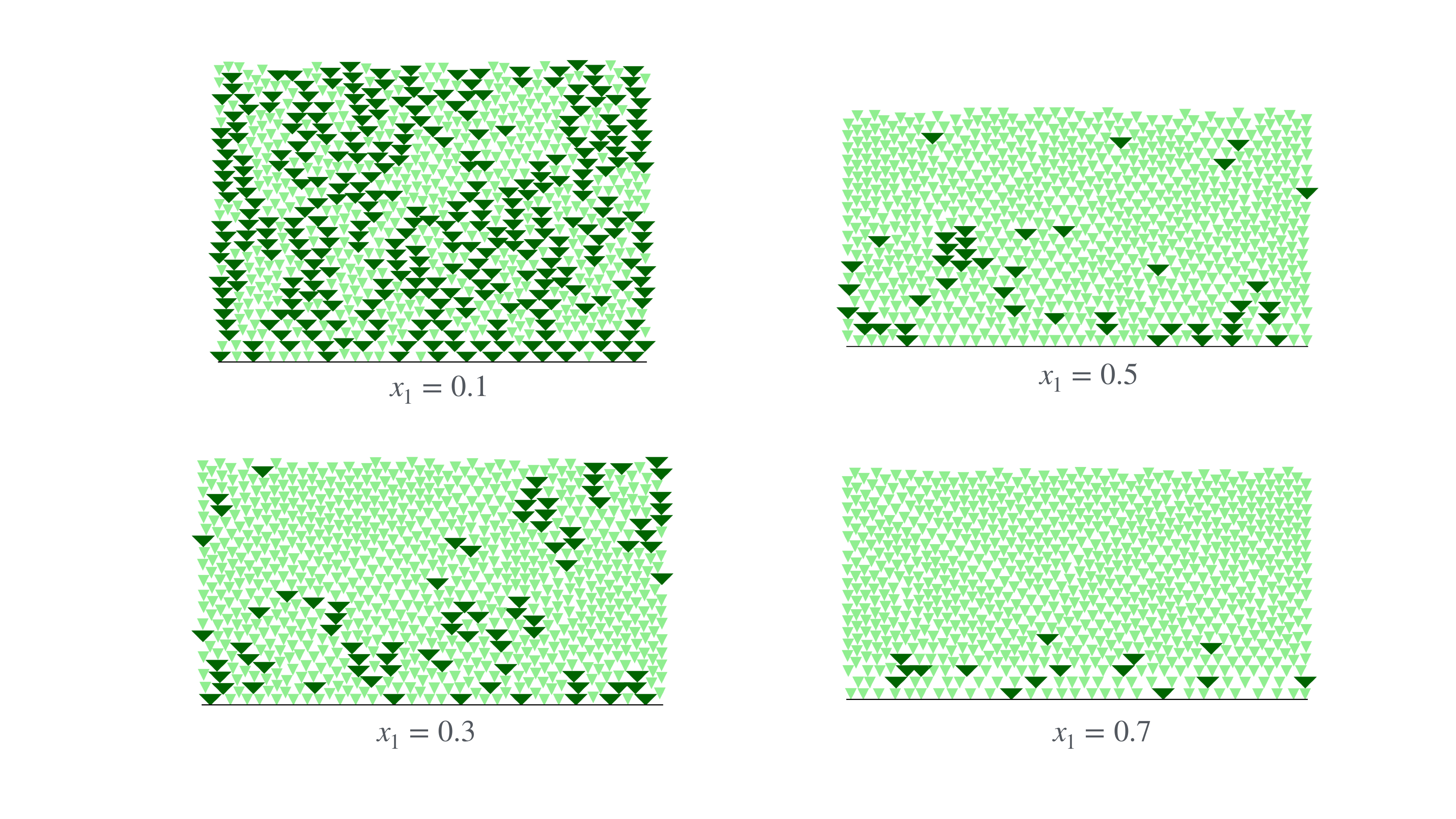}
   \caption{Event driven simulations of mixture M2 for different compositions. $L=40, ED = 40, k_p=500$.}
   \label{fig:M2}
   \end{center}
 \end{figure*}




\section{Conclusion}
We have investigated a one-dimensional random planting model with steric exclusion between neighboring plants using an event-driven simulation algorithm. For  monocultures, random planting at high rates results in high yields. Moreover, the limiting value seems to be equal that of an optimized  desynchronized regular planting process. In duocultures, despite the absence of attractive or repulsive interactions, a strong selectivity between species appears when either the size ratio the lifetime ratio deviate from unity. Our results provide a stepping stone to the investigation of higher dimensional systems.

\bibliography{}

\providecommand{\newblock}{}
\begin{thebibliography}{10}
\expandafter\ifx\csname doi\endcsname\relax
  \def\doi#1{{\tt #1}}\fi
\expandafter\ifx\csname doiprefix\endcsname\relax\def\doiprefix{DOI }\fi
\providecommand{\eprint}[2][]{\doi{#2}}

\bibitem{altieri2018agroecology}
Altieri M~A 2018 {\em Agroecology: the science of sustainable agriculture\/}
  (CrC press)

\bibitem{wezel2009agroecology}
Wezel A, Bellon S, Dor{\'e} T, Francis C, Vallod D and David C 2009 {\em
  Agronomy for sustainable development\/} {\bf 29} 503--515

\bibitem{de2011agroecology}
De~Schutter O {\em et~al.\/} 2011 {\em Report presented at the 16th session of
  the United Nations Human Rights Council [A/HRC/16/49]\/} {\bf 8}

\bibitem{food2021long}
Food I, Group E {\em et~al.\/} 2021 {\em IPES Food: Brussels, Belgium\/}

\bibitem{Orin}
Orin M accessed on the 09/01/2022 French intensive gardening: A retrospective

\bibitem{herve2016miraculous}
Herv{\'e}-Gruyer P and Herv{\'e}-Gruyer C 2016 {\em Miraculous abundance: One
  quarter acre, two French farmers, and enough food to feed the world\/}
  (Chelsea Green Publishing)

\bibitem{hanappe2016agroecology}
Hanappe P, Dunlop R, Maes A, Steels L and Duval N 2016 {\em Human
  Computation\/} {\bf 3} 225--233

\bibitem{raghavan2016computational}
Raghavan B, Nardi B, Lovell S~T, Norton J, Tomlinson B and Patterson D~J 2016
  Computational agroecology: Sustainable food ecosystem design {\em Proceedings
  of the 2016 CHI Conference Extended Abstracts on Human Factors in Computing
  Systems\/} pp 423--435

\bibitem{colliaux2022computational}
Colliaux D, Minchin J, Goelzer S and Hanappe P 2022 Computational agroecology:
  Should we bet the microfarm on it? {\em Eighth {Workshop} on {Computing}
  within {Limits} 2022\/} (LIMITS)

\bibitem{paut2020modelling}
Paut R, Sabatier R and Tchamitchian M 2020 {\em Agriculture, Ecosystems \&
  Environment\/} {\bf 288} 106711

\bibitem{colliaux2017lettucethink}
Colliaux D and Hanappe P 2017 Lettucethink: A open and versatile robotic
  platform for weeding and crop monitoring on microfarms {\em EFITA WCCA 2017
  Conference\/}

\bibitem{ditzler2022automating}
Ditzler L and Driessen C 2022 {\em Journal of Agricultural and Environmental
  Ethics\/} {\bf 35} 2

\bibitem{li2023productive}
Li C, Stomph T~J, Makowski D, Li H, Zhang C, Zhang F and van~der Werf W 2023
  {\em Proceedings of the National Academy of Sciences\/} {\bf 120} e2201886120

\bibitem{Paut2024}
Paut R, Garreau L, Ollivier G, Sabatier R and Tchamitchian M 2024 {\em
  Scientific Data\/} {\bf 11} 5

\bibitem{vandermeer1992ecology}
Vandermeer J~H 1992 {\em The ecology of intercropping\/} (Cambridge university
  press)

\bibitem{Ecormier-Nocca2019}
Ecormier-Nocca P, Memari P, Gain J and Cani M~P 2019 {\em Computer Graphics
  Forum\/} {\bf 38} 157--168

\bibitem{Tei1996}
Tei F, Aikman D and Scaife A 1996 {\em Annals of Botany\/} {\bf 78} 645--652

\bibitem{thornley2007mathematical}
Thornley J~H and France J 2007 {\em Mathematical models in agriculture:
  quantitative methods for the plant, animal and ecological sciences\/} (Cabi)

\bibitem{Garcia-Barrios2001}
García-Barrios L, Mayer-Foulkes D, Franco M, Urquijo-Vásquez G and
  Franco-Pérez J 2001 {\em Bulletin of Mathematical Biology\/} {\bf 63}
  507--526

\bibitem{Ford1975}
Ford E~D 1975 {\em Journal of Ecology\/} {\bf 63} 311--333

\bibitem{Huston1986}
Huston M 1986 {\em Ecology\/} {\bf 67} 265--269

\bibitem{colliaux2023models}
Colliaux D, Gravino P, Hanappe P, Talbot J and Viot P 2023 Models for the
  computational design of microfarms {\em International Conference on Complex
  Computational Ecosystems\/} (Springer) pp 121--132

\bibitem{Talbot2000}
Talbot J, Tarjus G, {Van Tassel} P and Viot P 2000 {\em Colloids and Surfaces
  A: Physicochemical and Engineering Aspects\/} {\bf 165} 287--324

\bibitem{Kubala2022}
Kubala P, Batys P, Barbasz J, Weroński P and Cieśla M 2022 {\em Advances in
  Colloid and Interface Science\/} {\bf 306} 102692

\bibitem{renyi1958one}
Renyi A 1958 {\em Publ. Math. Inst. Hungar. Acad. Sci\/} {\bf 3} 109--127

\bibitem{TalbotMixtures1989}
Talbot J and Schaaf P 1989 {\em Phys. Rev. A\/} {\bf 40}(1) 422--427

\bibitem{tarjus1991random}
Tarjus G and Talbot J 1991 {\em Journal of Physics A: Mathematical and
  General\/} {\bf 24} L913

\bibitem{Hassan2002}
Hassan M, Schmidt J, Blasius B and Kurths J 2002 {\em Physica A: Statistical
  Mechanics and its Applications\/} {\bf 315} 163--173

\bibitem{Wagaskar2020}
Wagaskar K~V, Late R, Banpurkar A~G, Limaye A~V and Shelke P~B 2020 {\em
  Journal of Statistical Physics\/} {\bf 181} 2191--2205

\bibitem{Boyer1994}
Boyer D, Talbot J, Tarjus G, Van~Tassel P and Viot P 1994 {\em Phys. Rev. E\/}
  {\bf 49}(6) 5525--5534

\bibitem{Dodds2002}
Dodds P~S and Weitz J~S 2002 {\em Physical Review E\/} {\bf 65} 056108

\bibitem{Andrienko1994}
Andrienko Y~A, Brilliantov N~V and Krapivsky P~L 1994 {\em Journal of
  Statistical Physics\/} {\bf 75} 507--523

\bibitem{Widom1966}
Widom B 1966 {\em The Journal of Chemical Physics\/} {\bf 44} 3888--3894

\bibitem{Torquato2006}
Torquato S, Uche O~U and Stillinger F~H 2006 {\em Phys. Rev. E\/} {\bf 74}(6)
  061308

\bibitem{Oleyar2007}
Oleyar C and Talbot J 2007 {\em Physica A: Statistical Mechanics and its
  Applications\/} {\bf 376} 27--37

\bibitem{Talbot2000b}
Talbot J, Tarjus G and Viot P 2000 {\em Phys. Rev. E\/} {\bf 61}(5) 5429--5438

\bibitem{Talbot1997}
Talbot J 1997 {\em AIChE Journal\/} {\bf 43} 2471--2478

\bibitem{Gillespie1977}
Gillespie D~T 1977 {\em The Journal of Physical Chemistry\/} {\bf 81}
  2340--2361

\bibitem{Rao2003}
Rao C~V and Arkin A~P 2003 {\em The Journal of Chemical Physics\/} {\bf 118}
  4999--5010

\bibitem{TV06}
Talbot J and Viot P 2006 {\em J. Phys. A: Math. Gen.\/} {\bf 39} 10947

\bibitem{talbot2024optimizing}
Talbot J 2024 unpublished

\end{thebibliography}

\end{document}